\documentclass[letterpaper, 10 pt, conference]{ieeeconf}  

\IEEEoverridecommandlockouts                              

\usepackage[table]{xcolor}%
\usepackage{arydshln} 
\usepackage{hyperref}

\usepackage{soul}
\usepackage{tabularx}
\usepackage{tabulary}
\usepackage{makecell}

\usepackage{multirow}
\usepackage{subcaption}
\usepackage{threeparttable}
\usepackage{colortbl}
\usepackage{graphicx}
\let\labelindent\relax
\usepackage[inline, shortlabels]{enumitem}

\usepackage[activate]{microtype}
\usepackage[natbib, bibencoding=utf8, citestyle=numeric, bibstyle=ieee, maxbibnames=999, maxcitenames=2, mincitenames=1, sortcites]{biblatex}
\bibliography{main.bib}

\newcommand{\eg}{e.\,g., }
\newcommand{\ie}{i.\,e., }

\usepackage[acronym, shortcuts, nohypertypes={acronym}]{glossaries}
\newacronym{AUC}{AUC}{area under the curve}
\newacronym{Bi-LSTM}{Bi-LSTM}{bidirectional long short-term memory neural network}
\newacronym{CCC}{CCC}{concordance correlation coefficient}
\newacronym{CNN}{CNN}{convolutional neural network}
\newacronym{ECG}{ECG}{electrocardiogram}
\newacronym{ML}{ML}{machine learning}
\newacronym{GBRT}{GBRT}{Gradient Boosted Regression Trees}
\newacronym{MAE}{MAE}{mean absolute error}
\newacronym{OM}{OM}{Other Runners Only Model}
\newacronym{RMSE}{RMSE}{root mean squared error}
\newacronym{RPE}{RPE}{received perception of exertion}
\newacronym{SGD}{SGD}{stochastic gradient descent}
\newacronym{sEMG}{sEMG}{surface-electromyography}
\newacronym{UAR}{UAR}{Unweighted Average Recall}
\newacronym{HR}{HR}{heart rate}
\newacronym{SVM}{SVM}{Support Vector Machine}
\newacronym{TDNN}{TDNN}{Time Delay Neural Network}

\makeatletter
\def\footnoterule{\kern-3\p@
  \hrule \@width 2in \kern 2.6\p@} 
\makeatother

\usepackage[capitalise, nameinlink, noabbrev]{cleveref}

\usepackage{graphics} 
\usepackage{amsmath} 
\usepackage{amssymb}  
\usepackage[capitalise]{cleveref}
\usepackage{booktabs}

\title{\LARGE \bf
Non-Invasive Suicide Risk Prediction Through Speech Analysis
}
\author{Shahin Amiriparian$^{1,4}$, Maurice Gerczuk$^{1}$, Justina Lutz$^{2}$, Wolfgang Strube$^{2}$ \\ Irina Papazova$^{2}$, Alkomiet Hasan$^{2,3}$, Alexander Kathan$^1$, Bj\"orn W. Schuller$^{1,4,5}$
\thanks{$^{1}$Chair of Health Informatics, The University Hospital of the Technical University of Munich, Germany}%
\thanks{$^{2}$Department of Psychiatry, Psychotherapy and Psychosomatics, Faculty of Medicine, University of Augsburg, District Hospital Augsburg, Germany}
\thanks{$^{3}$German Center for Mental Health, Munich, Germany}
\thanks{$^{4}$Munich Center for Machine Learning (MCML), Germany}
\thanks{$^{5}$Munich Data Science Institute (MDSI), Germany}
\thanks{$^{*}$ Corresponding author: {\tt\small shahin.amiriparian@tum.de}}}

\begin{document}

\maketitle
\thispagestyle{empty}
\pagestyle{empty}

\begin{abstract}

The delayed access to specialized psychiatric assessments and care for patients at risk of suicidal tendencies in emergency departments creates a notable gap in timely intervention, hindering the provision of adequate mental health support during critical situations. To address this, we present a non-invasive, speech-based approach for automatic suicide risk assessment. For our study, we collected a novel speech recording dataset from $20$ patients. We extract three sets of features, including wav2vec, interpretable speech and acoustic features, and deep learning-based spectral representations. We proceed by conducting a binary classification to assess suicide risk in a leave-one-subject-out fashion. Our most effective speech model achieves a balanced accuracy of $66.2\,\%$. Moreover, we show that integrating our speech model with a series of patients' metadata, such as the history of suicide attempts or access to firearms, improves the overall result. The metadata integration yields a balanced accuracy of $94.4\,\%$, marking an absolute improvement of $28.2\,\%$, demonstrating the efficacy of our proposed approaches for automatic suicide risk assessment in emergency medicine.

\end{abstract}

\section{INTRODUCTION}
\label{sec:intro}

With more than 700,000 deaths and an estimated 25 million non-fatal suicide attempts per year~\cite{walsh2017predicting}, suicide has become a widespread issue in our time. It can impact people of every age group and was the fourth most common cause of death for those aged 15-29\,years worldwide in 2019, being most prevalent in low- and middle-income countries\footnote{\url{https://www.who.int/news-room/fact-sheets/detail/suicide}}.
In addition to the loss of life, suicide also has a significant economic impact. For example, in the United States, the annual economic cost of suicide is estimated to exceed \$93 billion~\cite{shepard2016suicide}, including various financial effects such as healthcare expenses or lost productivity.
Moreover, the number of suicides is increasing, influenced by multiple factors, \eg difficult challenges in life, an economic burden, or past mental health issues~\cite{nordin2022suicidal}.

With over $90\,\%$ of people committing suicide suffering from at least one mental disorder, providing mental health support to individuals at risk of suicide is an essential aspect of prevention efforts~\cite{lonnqvist1995mental}. Studies, such as McClellan et al.~\cite{mcclellan2021impact}, further support this by demonstrating the efficacy of mental health treatment for reducing the number of suicide attempts.
However, the lack of sufficient clinical experts poses a major challenge in practice. Specialized psychiatric assessments and care for patients potentially at risk for suicidal tendencies are often not available in emergency departments or can only be provided after a very long delay. Hence, the clinical assessment of suicidality poses a particular challenge for doctors of various specialties in emergency care settings due to lacking psychiatric expertise at their disposal. 
As a result, the use of mental health services is very low, with no access to treatment options being reported as the main factor~\cite{hom2015evaluating}.

To this end, \ac{ML}-based approaches and their use for the diagnostic assessment of mental illness by medical professionals have been explored, offering a helpful addition in clinical emergency settings.
In previous work, primarily electronic health records (EHR) obtained from hospitals, meta-information such as character traits, or text-based data, \eg from Twitter, were employed.
Walsh et al.~\cite{walsh2017predicting} explored an EHR approach to predict suicidal risk and analyzed the risk over time from 7 days to 720 days before the attempted suicide. Their experiments showed that the accuracy increases the closer the date gets to the suicide attempt, achieving the best \ac{AUC} of $84\,\%$. Another study that utilized longitudinal EHR data for suicide risk prediction determined short- and long-term risk factors, leading to an \ac{AUC} of up to $86\,\%$~\cite{su2020machine}.

Macalli et al.~\cite{macalli2021machine} researched meta-information from college students. Their study applied random forest models using features such as substance use and sociodemographic or familial characteristics, resulting in an \ac{AUC} of $80\,\%$. Furthermore, they identified the four most important features (\ie suicidal thoughts, traits of anxiety, self-esteem, and depression symptoms).

In addition, previous studies explored text-based suicide risk prediction. Roy et al.~\cite{roy2020machine} trained several neural networks using posts from Twitter data related to psychological conditions such as burden, anxiety, or loneliness. Subsequently, they trained a random forest model for predicting suicidal ideation cases using the outputs of the previously trained networks, achieving an \ac{AUC} of $88\,\%$.
An extensive linguistic analysis was conducted by Homan et al.~\cite{homan2022linguistic}. Their findings show that suicidal thoughts contain more intensifiers and superlatives. In contrast, suicidal behaviors utilized more often pronouns, fewer modifiers and numerals, as well as more prepend words. 

However, only some works analyzed speech to predict suicidality. Building upon the observation that the characteristics of the patient's voice are more important than the content of what was spoken, Ozdas et al.~\cite{ozdas2004analysis} used Mel-cepstral coefficients in conjunction with Gaussian mixture models to predict near-term suicidal risk.
Cummins et al.~\cite{cummins2015review} further analyzed how common paralinguistic speech characteristics are affected by suicidality. Finally, \citet{Gerczuk24-EGS} explored gender-specific speech patterns of suicidal patients undergoing emergency admission.

With our work, we want to extend previous speech-based approaches, developing and evaluating a clinical decision-making assistant based on automated co-assessments of suicidality and voice recordings of affected patients.

\section{DATASET}
\label{sec:data}


We have collected a unique speech-based dataset from $20$ subjects undergoing emergency admission to the Department of Psychiatry at the District Hospital Augsburg, Germany. These recordings encompass three distinct types of speech activities: (i) reading of three short narratives\footnote{The three short stories were (1) ``Der Nordwind und Die Sonne'' (the story of `the north wind and the sun'), ``Gleich am Walde'' (the story of `right next to the forest'), and ``Der Hund und das Stück Fleisch'' (the story of `the dog and the piece of meat').} widely used within phonetics and speech pathology, each story was read twice; (ii) picture description where each patient describes a picture in their words; and (iii) isolated vowels production (/a:/, /e:/, /i:/, /o:/, /u:/). Each subject contributed eight recordings after admission to the emergency room (referred to as baseline), and some consented to record an additional eight upon discharge from hospital care. In our experiments, we restrict the analysis to the baseline samples to allow for a consistent validation procedure. The study doctor assessed the suicide risk level of each patient using the Likert scale [1 -- 6].



\begin{table}[]
    \caption{Dataset statistics: number of segments/utterances (\# utt.), mean duration ($\mu$), standard deviation ($\sigma$), minimum, maximum, and total duration ($\Sigma$ dur.) of audio recordings per sample type.}
\resizebox{\linewidth}{!}{%
\centering
    \begin{tabular}{lrrrrrr}
        \toprule
\textbf{Sample Type} & 
\textbf{$\#$ utt.} & 
\textbf{$\mu$} [s] & 
\textbf{$\sigma$} [s] & 
\textbf{min} [s] & 
\textbf{max} [s] & 
\textbf{$\Sigma$ dur.} [m] \\
\midrule
	    Pic. Desc. & 984 & 6.34 & 4.55 & 0.12 & 29.76 & 103.94  \\ 
	    Neut. Texts & 1261 & 5.62 & 3.24 & 0.32 & 19.71 &	118.05 \\
        Vowels & 95 & 0.30	& 0.28 & 0.02 & 1.32 & 0.47 \\
     \midrule
	    \textbf{Total} & \textbf{2340} & \textbf{5.70} & \textbf{3.97} &      \textbf{0.02} & \textbf{29.76} & \textbf{222.46} \\ 
     \bottomrule
    \end{tabular}
        }
    \label{tab:data-stats}
\end{table}


\section{METHODOLOGY}
\label{sec:method}
In our approach, we begin by preprocessing the audio recordings through volume normalization and speech segmentation (cf.~\Cref{ssec:segmentation}). Following this, we extract deep speech representations and interpretable audio features for each audio segment (cf.~\Cref{ssec:audio-features}) and fuse them with patients' metadata (cf.~\Cref{ssec:metadata}). Subsequently, we train machine learning models using the extracted features (cf.~\Cref{ssec:experimental_setup}).
In an iterative process, we augment the obtained features with metadata from each patient. This allows us to examine the complementarity between audio features and metadata.

\subsection{Segmentation}
\label{ssec:segmentation}

As a first step in the audio-based analysis of suicide risk, we split the speech recordings into smaller segments -- sentences for the picture descriptions and neutral texts and individual vowels for the vowel production task. This process is automated in whisperX~\cite{bain2022whisperx} by force-aligning transcriptions generated by Whisper~\cite{Radford23-RSR} to the outputs of a wav2vec phoneme recognition model. We then cut the audio samples at the detected sentence or vowel boundaries. Statistics about the resulting audio segments are provided in~\Cref{tab:data-stats}.

\subsection{Audio Features}
\label{ssec:audio-features}

We extract a range of audio features from the segmented speech recordings, representing different state-of-the-art and traditional paradigms of automatic paralinguistic analysis, and evaluate their efficacy for suicide risk assessment. 

\subsubsection{Speech and Acoustic Functionals}
We utilize openSMILE~\cite{Eyben10-OTM} to extract two sets of features -- a comprehensive set of 6,373 functionals (\textsc{ComParE}) and \textsc{eGeMAPS}, a smaller 
selection of features (88 dimensional) crafted for speech emotion recognition. 

\subsubsection{Deep Audio Representations}
We enhance our feature set by incorporating \textsc{DeepSpectrum} features~\cite{Amiriparian17-SSC}, proven to be effective across various speech-based health-related and affective computing tasks~\cite{gerczuk2023noise, amiriparian2021ai, amiriparian2023muse}. Additionally, we include embeddings from three distinct pretrained audio Transformers: wav2vec-large~\cite{baevski2020wav2vec} and two fine-tuned versions, one on German~\cite{grosman2021xlsr53-large-german} and the other on the MSP-Podcast~\cite{lotfian2017building} specifically fine-tuned for speech emotion recognition~\cite{Wagner23-DOT}. We chose these latter two Transformers to explore their effectiveness on German data and to assess whether a model fine-tuned for emotion tasks can capture relevant patterns in the speech of individuals at risk of suicide.
For \textsc{DeepSpectrum} features, we first transfer audio signals into Mel-spectrogram plots with $128$ Mels and forward them through a \textsc{DenseNet201} architecture (weights pre-trained on ImageNet). Afterward, we extract the activations of the penultimate fully connected, resulting in a $1920$-dimensional feature set. 
For all wav2vec features, we extract $1024$-dimensional embeddings by passing the speech recordings through the pretrained wav2vec models. These embeddings are obtained by averaging the resulting representations from the final layer of the models.


\subsection{Metadata Fusion}
\label{ssec:metadata}
We further combine the audio features with a set of metadata collected from each patient to assess whether they can complement each other and enhance the overall results. The meta-features include demographic details (age, gender, height, body weight), history of suicide attempts, access to firearms or potentially lethal medication, levels of hopelessness, experiences of sexual abuse or trauma, exposure to stressful living conditions, instances of substance abuse, presence of manic episodes, occurrences of non-suicidal self-injury (NSSI), and scores from the Beck-Depressions-Inventar (BDI).

\subsection{Experimental Setup}
\label{ssec:experimental_setup}
We conduct binary classification of suicide risk based on assessments from the study doctor. Both physicians rate suicidality on a scale ranging from 1 (not suicidal) to 6 (acute suicide attempt). These ratings are then categorized into classes representing low suicidality (1-4) and high risk of suicide (5 and 6), forming the targets for our classification models. We evaluate these models using a leave-one-subject-out (LOSO) validation scheme. Classification is performed by fitting linear \ac{SVM} models on the extracted audio features and vectorized metadata, with categorical features like gender encoded using one-hot encoding. The data is normalized to zero mean and unit standard deviation based on each fold's training partition. We optimize the \ac{SVM}'s regularization constant on a logarithmic scale from $1$ to $10^{-7}$ via an inner 5-fold cross-validation. The evaluation metric is balanced accuracy, which accounts for class imbalances by considering the average accuracy of each class.



\section{RESULTS}
\label{sec:results}
\Cref{tab:results-audio} shows results measured in balanced accuracy achieved by relying only on speech recordings. We investigate the influence of speech content on classifier performance, training and evaluating the \ac{SVM} models on picture descriptions, neutral texts, vowel production, or all of the aforementioned. Generally, the best results are achieved when basing the machine learning analysis on the recordings of vowels. Here, a model trained on the minimalistic eGeMAPS set of audio functionals achieves a balanced accuracy of $66.2\,\%$. For the same type of speech recording, a purely spectral audio representation in the form of \textsc{DeepSpectrum} still performs over the chance level at $58.2\,\%$. From the pre-trained wav2vec models, the original large XSLR variant's features perform best, with neither German language nor emotion recognition fine-tuning yielding any benefits -- the latter failing to exceed the chance level in any experimental configuration. Intuitively, wav2vec's advantage of implicitly modeling linguistics~\cite{Wagner23-DOT} does not apply to recordings of single vowels, whereas spectral information can be exploited more easily in this case. 

\begin{figure}[ht!]
      \centering
    \includegraphics[width=.8\columnwidth]{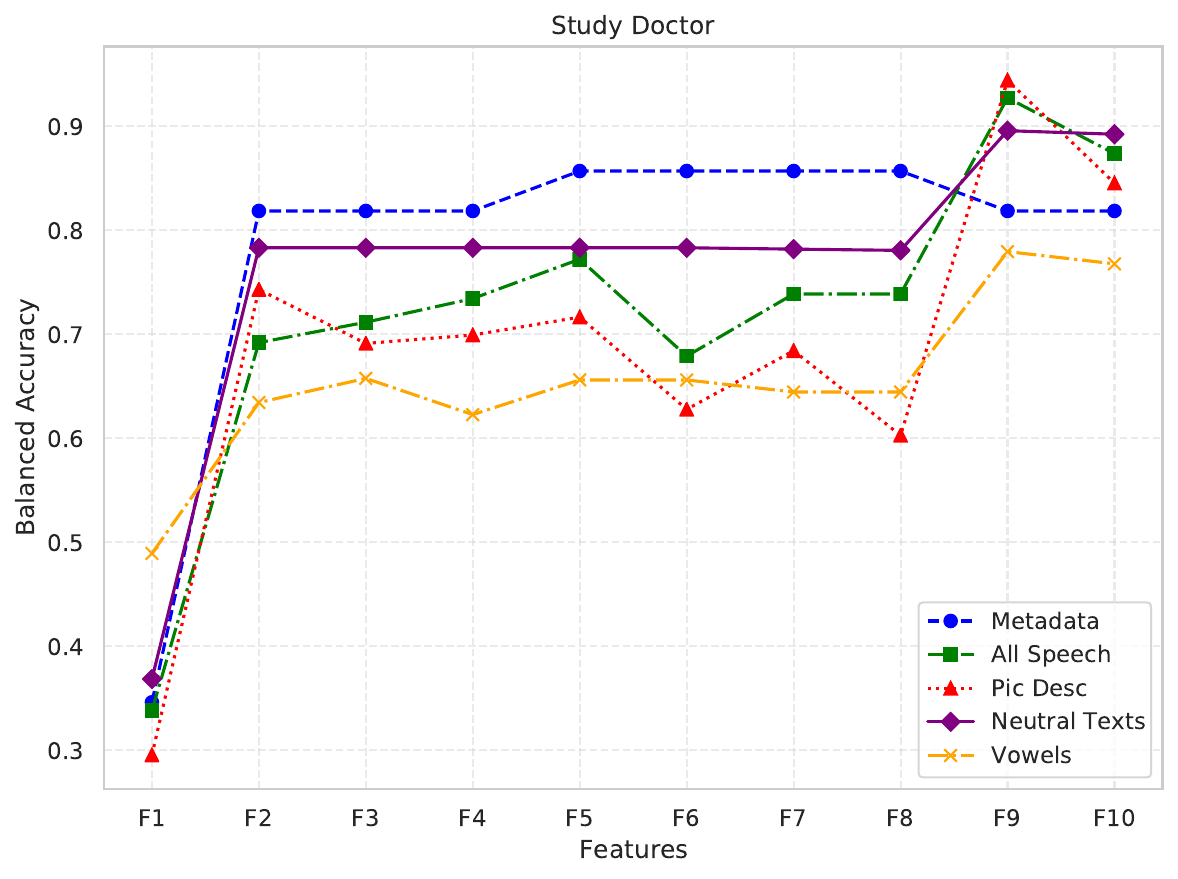}
    \caption{Comparison of models trained solely with metadata (blue) and models with metadata fused with features from different speech recordings: all speech data (green), speech data exclusively from picture descriptions (red), reading neutral texts (purple), and distinct vowels (orange). Feature types F1--F10: cf.~\Cref{tab:metadata-fusion}}
\label{fig:results-study-doctor}
\end{figure}

The results for automatically assessing suicide risk based on only metadata (\Cref{tab:metadata-fusion}) show that the most discriminative feature can be found with a patient's history of suicide attempts. Including this information in addition to baseline demographic information raises balanced accuracy from below chance level to $81.9\,\%$. Without fusion, only a moderate performance boost can be achieved by including more meta-information. Finally, we analyze the effects of adding eGeMAPS features to the metadata before classification (additionally visualized in~\Cref{fig:results-study-doctor}). Contrary to the audio-only results, adding features extracted from recordings of vowel productions generally leads to worse results than adding the same features extracted from picture descriptions and neutral texts. Neutral texts combined with metadata lead to the most consistent results, improving balanced accuracy to $89.6\,\%$ when including all but the BDI score. The overall best result, however, can be found when fusing audio representations extracted from recordings of the picture descriptions, reaching $94.4\,\%$ balanced accuracy. During the study, this part of the recording was found to be difficult and often strenuous for the participants. We hypothesize that this type of speech is more heavily impacted by the subjects's psychological state as it requires a substantially higher cognitive effort.

\begin{table}[]
    \resizebox{\columnwidth}{!}{
    \centering
    \begin{tabular}{lrrrr}
\toprule
 & \multicolumn{1}{r}{All Speech} & \multicolumn{1}{r}{Pic. Desc.} & \multicolumn{1}{r}{Neut. Texts} & \multicolumn{1}{r}{Vowels} \\
\midrule
\textsc{ComParE} & 46.3 & 52.3 & 48.8 & 64.0 \\
\textsc{eGeMAPS} & 50.4 & 45.2 & 58.8 & \cellcolor{gray!40}\textbf{66.2} \\
\hdashline
\textsc{DeepSpectrum} & 47.3 & 40.2 & 40.1 & 58.2 \\
\hdashline
w2v-emotion & 48.4 & 51.0 & 52.0 & 48.5 \\
w2v-large & \textbf{54.9} & 47.4 & \textbf{60.5} & 55.7 \\
w2v-large-german & 48.9 & 39.7 & 53.1 & 55.2 \\
\bottomrule
\end{tabular}
}
    \caption{Results for audio-based suicide risk classification utilizing target labels for suicide risk assessed by the study doctor, with performance evaluated in terms of balanced accuracy. The best result for each speech category is boldfaced, and the best overall result is marked with gray shading.}
    \label{tab:results-audio}
\end{table}

\begin{table*}[ht!]
    \centering
    \begin{tabular}{lrrrrr}
\toprule
 \multicolumn{2}{c}{Metadata} & \multicolumn{4}{c}{Fusion of Metadata with Speech} \\
\cmidrule(lr){1-2} \cmidrule(lr){3-6}
  & \multicolumn{1}{r}{Only Metadata} & \multicolumn{1}{r}{All Speech} & \multicolumn{1}{r}{Pic. Desc.} & \multicolumn{1}{r}{Neut. Texts} & \multicolumn{1}{r}{Vowels} \\
  \cmidrule(lr){2-2} \cmidrule(lr){3-3} \cmidrule(lr){4-4} \cmidrule(lr){5-5} \cmidrule(lr){6-6}
 Demographics (F1) & 34.6 & 33.8 & 29.5 & 36.9 & 48.9 \\
 F1 + Suicide Attempts (F2) & 81.9 & 69.2 & 74.3 & 78.3 & 63.4 \\
 F2 + Firearms or Potentially Lethal Medication (F3) & 81.9 & 71.2 & 69.1 & 78.3 & 65.8 \\
 F3 + Hopelessness (F4) & 81.9 & 73.4 & 69.9 & 78.3 & 62.3 \\
 F4 + Sexual Abuse/Trauma (F5) & 85.7 & 77.2 & 71.7 & 78.3 & 65.6 \\
 F5 + Stress Situation (F6) & 85.7 & 67.9 & 62.8 & 78.3 & 65.6 \\
 F6 + Substance Abuse (F7) & 85.7 & 73.9 & 68.4 & 78.2 & 64.5 \\
 F7 + Mania (F8) & 85.7 & 73.9 & 60.3 & 78.1 & 64.5 \\
 F8 + NSSI (F9) & 81.9 & 92.7 & \cellcolor{gray!40}\textbf{94.4} & 89.6 & 78.0 \\
 F9 + BDI (F10) & 81.9 & 87.4 & 84.5 & 89.2 & 76.8 \\
\bottomrule
\end{tabular}
    \caption{Metadata and Fusion Results: The first column shows results of models trained with only metadata, while subsequent columns show results of fusing metadata with the top speech model. F2 to F10 are generated by incrementally adding metadata. The best result is shaded gray.}
    \label{tab:metadata-fusion}
\end{table*}

\section{CONCLUSIONS AND FUTURE WORK}

We have introduced a non-invasive, speech-based approach for automatic suicide risk assessment. Leveraging a novel dataset of speech recordings from $20$ patients, we extracted six sets of audio representations, including interpretable speech and acoustic features, \textsc{DeepSpectrum} features, and embeddings from pretrained audio Transformers. We showed that our model can recognize the degree of suicidality (high/low) from speech with a balanced accuracy of $66.2\,\%$, and when it is fused with patients' metadata, a balanced accuracy of up to $94.4\,\%$ can be achieved.

Future work involves expanding the dataset to a larger, more diverse cohort and improving model generalization. Additional metadata like social media activity or lifestyle habits could enhance accuracy. Incorporating longitudinal data to track speech pattern changes may enable dynamic risk assessment of suicidal tendencies~\cite{gerczuk2023zero}. Finally, using advanced affective computing models like ExHuBERT\footnote{\href{https://huggingface.co/amiriparian/ExHuBERT}{https://huggingface.co/amiriparian/ExHuBERT}}~\cite{Amiriparian24-EEH} could offer deeper emotional analysis, further refining risk prediction.



\section*{\refname}
\printbibliography[heading=none]

\end{document}